\documentstyle[11pt,twoside,paspconf]{article}

\begin{document}

\title{The DDO IVC Distance Project}

\author{Michael D. Gladders, Tracy E. Clarke, Christopher R. Burns,
  Allen Attard, Michael P. Casey, Devon Hamilton, Gabriela
  Mall\'{e}n-Ornelas, Jennifer L. Karr, Sara M. Poirier, Marcin Sawicki,
  Felipe Barrientos, Wayne Barkhouse, Mark Brodwin, Jason Clark,
  Rosemary McNaughton, Marcelo Ruetalo-Pacheco and Stefan W.
Mochnacki} \affil{University of Toronto \& The David Dunlap
  Observatory}

\begin{abstract}
We present the first set of distance limits from
the David Dunlap Observatory Intermediate Velocity Cloud (DDO IVC)
distance project. Such distance measures are crucial to understanding
the origins and dynamics of IVCs, as the distances set most of the
basic physical parameters for the clouds. Currently there are very few
IVCs with reliably known distances. This paper describes in some
detail the basic
techniques used to measure distances, with particular emphasis on the
the analysis of interstellar absorption line data, which forms the
basis of our distance determinations. As an example, we provide a
detailed description of our distance determination for the Draco
Cloud. Preliminary distance limits for a total of eleven clouds are
provided.
\end{abstract}


\section{Introduction}
Since their discovery in 1963 by Muller, Ooort \& Raimond, the high- and
intermediate-velocity clouds (HVCs and IVCs respectively) have been
somewhat of a mystery. Central to our lack of understanding of these
discrete clouds is our lack of knowledge of their distances. The
distances set many of the basic physical parameters of the clouds (see
Table 1) and are critical in interpreting origin and evolution
models for these objects (see Wakker and van Woerden 1997 - hereafter
WvW97 - for a review
of models). Furthermore, both the IVCs and HVCs show
significant correlations with Galactic x-ray emission (e.g. Snowden
{\it et al.} 1991, Kerp {\it et al.} 1998, Gladders {\it et al.} 1998) and knowledge of the
distances to these clouds helps to constrain models of the x-ray
emitting gas in the Galaxy. The HVCs have received significant observational
attention in recent years, and distance limits have now been set to
several large HVC complexes (WvW97). A similar concerted effort
has not occurred in determining distances for the IVCs, and distance
limits have been set for only a few IVCs.

\begin{table}
\caption{The dependence of some physical parameters of IVCs on the
  cloud distance (WvW97).}
\begin{center}
\begin{tabular}{lc}
\tableline\tableline
Column Density & d$^{0}$\\
Density& d$^{-1}$\\
Linear Size & d$^{+1}$\\
Mass& d$^{+2}$\\
Metallicity& d$^{0}$\\
Pressure& d$^{-1}$\\
\tableline\tableline
\end{tabular}
\end{center}
\end{table}
\paragraph{}
Several methods have been used for determining HVC and IVC distances.
The most direct and convincing of these is the absorption line method
(e.g. Lilienthal {\it et al.} 1991, Benjamin {\it et al.} 1996). The
basic technique is to obtain relatively high resolution
spectra of probes (typically, early-type stars) along the
line-of-sight to the target cloud, searching for interstellar
absorption from appropriate metallic species at the cloud's systemic
velocity. Any star showing such absorption can then be considered to
be behind the cloud, and any star not showing such absorption (modulo
covering factor effects - see Burns {\it et al.} 1999) to be in front of the cloud.
Distance estimates to the bracketing stars, typically determined from
spectroscopic parallax, then provide a distance bracket for the cloud.
The primary drawback to this approach is that it requires a large
investment of telescope time, as the search for stars showing the
expected absorption may require data on numerous potential background
sources. Moreover, in the absence of detected absorption, the
interpretation of non-detections of absorption is somewhat ambiguous
(WvW97).
\paragraph{}
An ongoing project has been started at the David Dunlap Observatory
(DDO) to address the lack of distance measures to IVCs via the
absorption line technique. The primary goal of this project is to
obtain both upper and lower distance estimates for a significant
sample of IVCs. The potential astrophysical applications of this work
are described elsewhere (Clarke {\it et al.} 1998); in this paper we
concentrate on a thorough description of our methodology, and give
some first results. \S 2 describes our target selection process,
observations and data reduction techniques. \S 3 describes our
analysis procedures, both for the detection/non-detection of
cloud-induced interstellar absorption and our spectroscopic
classification and resulting distance measures for useful bracketing
stars. \S 4 gives preliminary
distances for a total of 11 IVCs.

\section{Target Selection and Observations}
\subsection{Cloud Targets}
The primary cloud sample we have been observing is drawn from the IVC
sample of Heiles, Reach and Koo (1988;  hereafter HRK88). These clouds
were selected primarily on the basis of their $IRAS$ 100$\mu$
emission, and form a relatively homogeneous set of isolated
interstellar clouds. Each cloud is $\sim1^{\circ}$ in size, and most
are at high galactic latitudes.  From the HRK88 sample of 26 clouds,
we have selected a subset of 16 clouds, all readily observed from the DDO.
Of these, we have eliminated G137.3+53.9 and G135.3+54.5 (the UMa IVC) from our
sample, as these two poorly separated clouds already have a measured
distance (Benjamin {\it et al.} 1996). We have retained the nearby cloud
G135.5+51.3 (considered by Benjamin {\it et al.} to be part of the
UMa Cloud) as examination of the HI atlas of Hartmann and Burton
(1997) indicates that this cloud may be a separate object. Conversely,
we have combined three pairs of clouds from the HRK88 sample, due both
to their apparent lack of separation on the sky (in both HI and $IRAS$
100$\mu$ emission) and their lack of separation in velocity space(HRK88).
The final cloud sample thus contains eleven clouds (Table 2).
\subsection{Stellar Targets}
The primary catalog of potential stellar probes we have used is the
Tycho catalog (Hog {\it et al.} 1997), occasionally supplemented by the
USNO catalog. The Tycho catalog provides positions and $B-V$
colours for over one million stars covering the entire sky. The sample is
complete to V$\sim$11, and contains stars to V$\sim$12. At the
completeness limit, the accuracy of the $B-V$ colours from the catalog
is $\sim$0.1, more than sufficient to give a rough indication of the
spectral class of each star. Some knowledge of the spectral class of
potential target stars is required, as this can be used to provide a
rough estimate of the distance to each star. Furthermore, the
interstellar absorption line method requires stars of relatively early
spectral type, as these stars provide a relatively clean stellar
continuum against which interstellar absorption is readily
identified. 
\paragraph{}
Our stellar target selection process for a given cloud proceeds as
follows. First, we delineate a bounding contour for the cloud in HI,
using the HI maps of Hartmann and Burton
(1997). This boundary is selected from an $\ell$,$b$ map of a velocity
slice  $\pm$5 km~s$^{-1}$ about the cloud's systemic velocity as
given by HRK88. The boundary is set at a flux level which clearly delineates
the cloud relative to the background. We then select all stars from
the Tycho catalog which lie interior to this boundary, and which
have a blue colour ($B-V<$0.4). Such stars are likely to be early type
stars with spectral classes earlier than F5, although the large colour
errors in the Tycho catalog at the flux limits of the sample do
cause some contamination by later-type stars. We then produce an HI
spectrum and a nominal distance estimate for each star. The distance
estimate is made by assigning a spectral class based on colour and
assuming the star is a main-sequence dwarf, thus assigning an
absolute magnitude to each star. This distance estimate may be quite
uncertain, due to large colour errors (see \S 4 for more details). 
A subset of these potential target stars is then selected on the
basis of distance and HI column due to the cloud. This
subset is then placed in our observing queue and observed as
conditions dictate.
\subsection{Observations}
All the observations undertaken for this project have been made with
the DDO 1.88m telescope + Cassegrain spectrograph. A total of 84
nights have been assigned to this project since May 1997, of which 59
nights have produced at least some useful spectroscopic observations.
These spectroscopic observations are of two flavors - moderate
resolution spectra at the Na~I doublet (to search for interstellar
absorption) and lower resolution 'classification spectra' (to refine
our stellar spectral classifications). Each type of observation and
any supplementary observations of each type are described in detail
below.
\subsubsection{Interstellar Absorption}
The interstellar absorption spectra have all been acquired using the DDO
Cassegrain spectrograph at its finest possible resolution,
corresponding to a velocity resolution of 22 km~s$^{-1}$ at the Na~I
doublet. This resolution is coarser than used by other authors
(e.g. Benjamin {\it et al.} 1996), but is sufficient to detect (though
not resolve) the absorption features due to IVCs. We have observed
stars to a limiting magnitude of $V\sim12.5$; on such targets we
achieve a S/N of ~50 in 4 hours. Target stars are observed as
conditions dictate, with the faintest targets reserved for the best
conditions. We also observe several telluric standards repeatedly
during each night. Numerous telluric standards have been
observed in the
course of this project, and we have found only two stars (HD177724 and
HD120315) which are not polluted by low-velocity interstellar Na~I
absorption. These two stars are the primary telluric standards used
for all data (see Figure 1). We have also obtained spectra of numerous
spectral classification standards with the Na~I setup, to provide the
necessary input into our analysis  techniques (see \S3.1).
\begin{figure}[!htb]
\plotfiddle{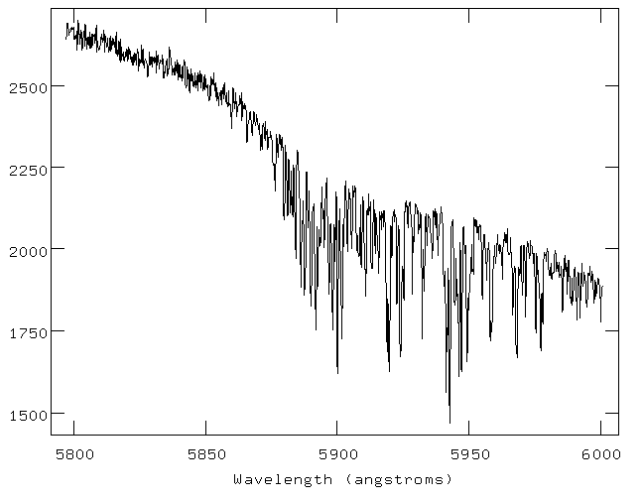}{75truemm}{0}{45}{45}{-150}{-80}
\caption{A typical telluric spectrum of HD177724, in arbitrary
  intensity units. As mentioned in the
  text, only HD177724 and HD120315 have been used as telluric
  standards, as all other tellurics observed showed significant
  low-velocity Na~I absorption. }
\end{figure}

\subsubsection{Spectral Classification}
The spectral classification setup used covers a spectral range of
3800-4400\AA, with a resolution of 2.5\AA. Classification spectra
have been acquired after the sodium observations, for only a subset of
all stars observed at the Na~I doublet. We have not taken
classification spectra of target stars prior to the acquisition of
sodium spectra because the spectrograph used is very inefficient in
the blue, mandating roughly equal total integration times for both
spectra. This makes the pre-selection of distant targets through
spectral classification no more efficient than direct observations of
likely targets identified by colour. We have also acquired spectra of a
suite of spectral classification standards (taken from Garcia 1989),
ranging in spectral type from B0 to G5, and spanning luminosity
classes I-V.

\begin{figure}[!htb]
\plotfiddle{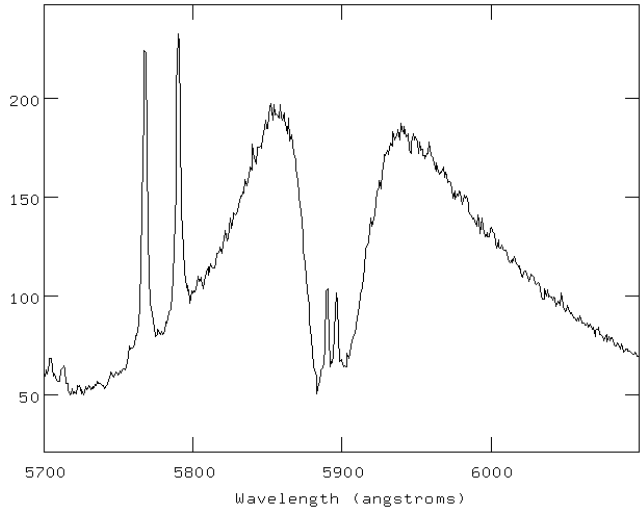}{75truemm}{0}{45}{45}{-150}{-80}
\caption{The sky spectrum at DDO from 5700-6100\AA, in arbitrary
  intensity units. Note the broad emission from high-pressure sodium
  lamps, which is strongly self absorbed at the doublet wavelengths.
  The most troublesome features of the sky spectra are the sharp
  emission lines due to low pressure sodium lamps. Correct subtraction
  of these lines is essential. }
\end{figure}

\subsection{Data Reduction Techniques}
The reduction of both types of spectral data is accomplished via a
data-reduction pipeline, implemented under IRAF \footnote{IRAF is
  distributed by the National Optical Astronomy Observatories, which
  is operated by the Association of Universities for Research in
  Astronomy, Inc., under contract to the National Science Foundation.
  }. The data reduction follows relatively standard longslit spectral
reduction procedures, and will not be described in great detail here.
Two things should be noted, however. First, the data are usually
processed to telluric corrected, wavelength calibrated,
heliocentrically corrected 1-D spectra within 24 hours of acquisition.
This rapid processing allows us to use recent observations of a given
cloud to guide the selection of other target stars. Second, great care
is given to the subtraction of sky-lines in the Na~I doublet spectra,
as DDO's close proximity to Toronto makes it a highly light-polluted
site. Luckily, most of the sodium lights near DDO are high pressure
lamps, which are strongly self absorbed at the doublet wavelengths,
with only a small contribution from low-pressure lamps (see Figure 2).

\section{Analysis Techniques}
The analysis of the spectroscopic data can be roughly divided into
three parts, each of which is described in some detail below. The first
is the identification of detections and non-detections of interstellar absorption
in the Na~I spectra. The second is the assignment of each star to
either the foreground or background relative to the cloud. Though the
background identifications are simple  once an absorption detection is
made, the unambiguous identification of a star as a foreground
object is non-trivial and subject to several uncertainties. The
third step in the analysis is to assign a distance to each foreground
and background star via spectroscopic parallax, and hence produce a
distance bracket to the target cloud.
\subsection{Detections of Interstellar Absorption}
The correct identification of absorption due to the target IVC is one
of the critical steps in our analysis. The most significant sources of
confusion are intrinsic stellar lines mis-identified as interstellar
features. Unlike other authors (e.g. Benjamin {\it et al.} 1996), we
have used somewhat later type stars in our target sample, and so must
be more concerned about possible contamination by stellar sodium
lines. Furthermore, our spectra are not of sufficient resolution to
resolve the interstellar absorption features expected from the target
IVCs (HRK88), and so we cannot rely upon the line profiles as an
indicator of the origin of observed sodium absorption lines. However,
the use of only moderate dispersion does provide a significant gain in
the wavelength sampling of the underlying stellar spectrum, and we use
this fact to motivate our analysis. For any sodium spectrum, the
analysis proceeds as follows.
\paragraph{}
First, we make a rough estimate of the spectral class of the star
using the sodium spectrum itself (or, if available, a classification
spectrum - see \S3.3). We then select a matching standard star from our
database of standards and measure the equivalent width and central
wavelength of all significant features in this standard spectrum.
These measurements are used to construct a synthetic spectrum; the
synthetic spectrum has a flat continuum, with sharp (FWHM $<<$
spectral resolution in real spectra) absorption features with
equivalent widths consistent with the standard at V$_{LSR}=0$ (Figure
3).
\begin{figure}[!htb]
\plotfiddle{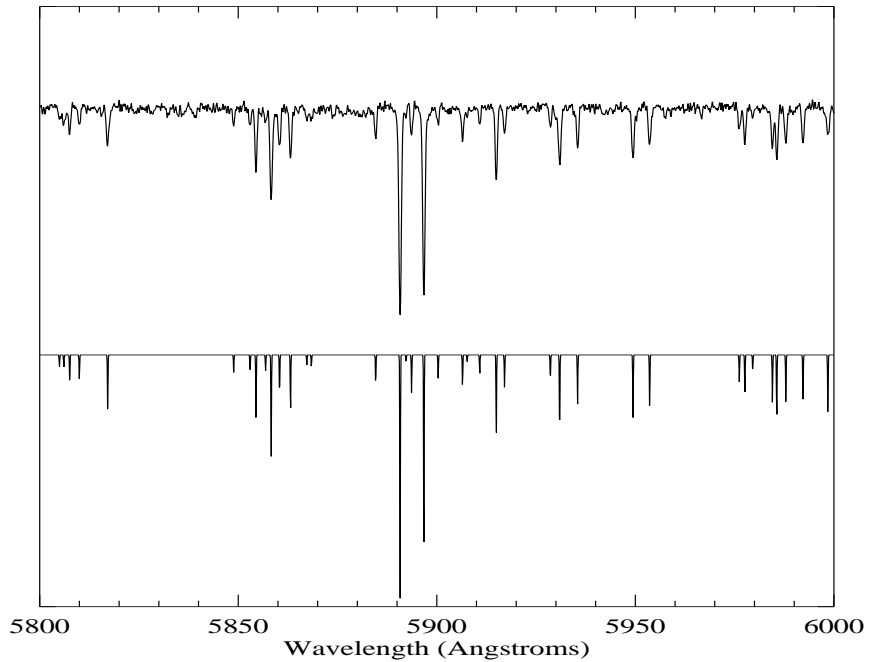}{95truemm}{0}{95}{65}{-210}{0}
\caption{An example of a synthetic spectrum constructed from a
  standard star spectrum, in arbitrary intensity units. The input
  spectrum, for the F6V star HD173667, is shown at the top, and the
  resulting synthetic spectrum, not at the same scale, is shown below it.}
\end{figure}

\paragraph{}
Next, we cross-correlate this synthetic spectrum with the target
spectrum in two regions. The first cross-correlation is performed only
in the 'stellar' region of the spectrum. This is essentially all of
the input spectrum {\it except} the Na~I doublet region. The resulting
cross-correlation gives the radial velocity and width of the stellar
features. The second cross-correlation is performed {\it only} in the
Na~I doublet region, and gives the radial velocity and width of the
stellar features {\it and} interstellar features. A simple comparison
of the two cross-correlations then determines which, if any, Na~I
features can be attributed to interstellar absorption. Figure 4
details this process for a star for which the interstellar absorption
component attributed to the intervening IVC is clearly distinguished,
despite serious contamination by stellar sodium features at
essentially the same radial velocity.

\begin{figure}[!htb]
\plotfiddle{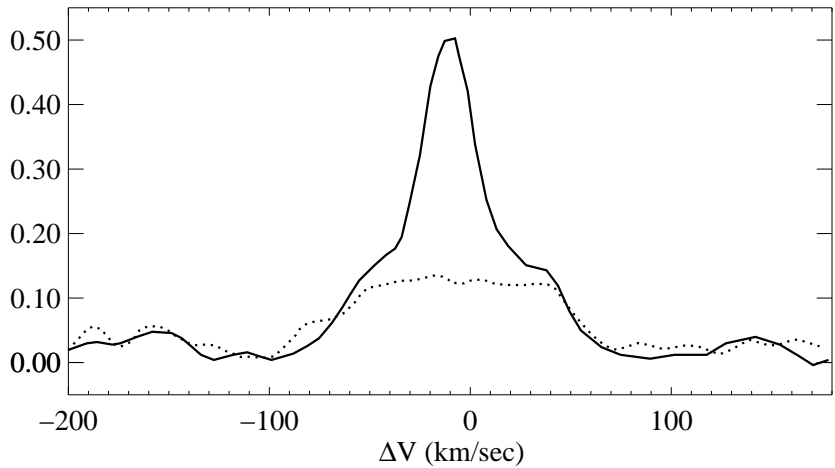}{75truemm}{0}{75}{75}{-210}{0}
\caption{The cross-correlation results for TYC~4152~484~1. The dotted
  line shows the cross-correlation for the 'stellar region', and the
  solid line for the sodium region. Note that the broad stellar peak
  is represented in the sodium peak, as expected, but that the
  additional sharp component in the sodium cross-correlation shows no
  corresponding stellar feature. This peak, at V$_{LSR}\sim-12$
  km~s$^{-1}$, is due to the IVC G139.6+47.6.}
\end{figure}

\subsection{Foreground and Background Stars}
The identification of detections and non-detections is only the first
step in determining the placement of stars in relation to the target
IVC. Critical to the distance determination process is the correlation
between non-detections (at a given significance) and foreground
objects. There are several reasons why this is non-trivial. First,
stellar probes provide a pencil beam for the detection of Na~I
absorption, which is then compared to a much larger beam from HI data.
At smaller scales than the HI data the clouds may have a covering
factor which is significantly less than unity, which could produce
non-detections for stars which are {\it behind} the target cloud.
Second, in the absence of detections of Na~I absorption due to the
cloud, the metallicity of the cloud remains uncertain, and so the
{\it expected} Na~I equivalent width {\it given} a HI column density
is uncertain, quite apart from small-scale covering factor effects.
The problem of metallicity is easily addressed; all non-detections are
to a degree ambiguous without at least one detection to set the HI/Na~I
ratio. Of course, one can extrapolate the expected HI/Na~I ratio from
other clouds in the same sample, assuming the ratio is constant. In
the absence of further data, this is a reasonable approach. Notably,
the cloud sample we are observing is large enough to directly test
this assumption. The problem of small-scale structure is more
difficult to surmount. One possible approach is to compare the
relationship between detections of Na~I absorption and distance for a
sample of stars towards a single cloud in which more than a single
detection and non-detection have been made. For one cloud in our sample
(G139.6+47.6 + G141.1+48.0) we have observed a total of 20 stars, and
made 8 detections. A preliminary analysis of these data indicate a
one-to-one correspondence between detection of absorption and distance,
in that the eight stars showing absorption are also the eight most
distant. A full treatment of these data will be presented elsewhere
(Burns {\it et al.} 1999), but the preliminary indication is that the small
scale covering factor is near unity. 
\paragraph{} 
Armed with this conclusion, we now proceed as follows in determining
the placement of target stars relative to a given cloud. First, the
detected absorption is used to set the HI/Na~I ratio for the cloud. In
the absence of detected absorption, the mean observed for other clouds
is used. We then extract HI spectra towards each target star from the HI
data of Hartmann \& Burton (1997) and measure the HI column density
due to the cloud along each line of sight. These data are then used to
predict an expected Na~I equivalent width for each star, assuming that
the stars are behind the target cloud. The predicted Na~I absorption
can then be compared to the observed spectra, which are nominally
non-detections of absorption. In 
each case, the expected absorption must be compared to the sensitivity
of each spectrum; in cases where the expected absorption
may be lost in the noise, the stars must either be discarded from the
sample, or re-observed to resolve the ambiguity. Only stars in which
the expected absorption is clearly un-detected can be taken as
foreground stars. Figure 5 demonstrates this analysis for four stars
towards the Draco Cloud (G90.0+38.8 + G94.8+37.6)
\begin{figure}[!htb]
\plotfiddle{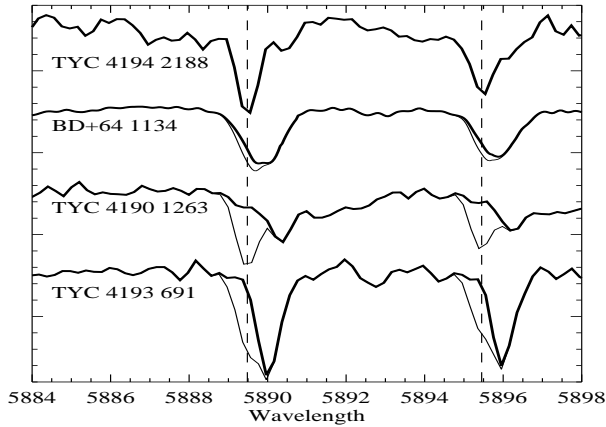}{65truemm}{0}{95}{65}{-210}{-30}
\caption{Na~I spectra for four stars towards the Draco Cloud, from
  Gladders {et al.} (1998). Measured spectra are shown as thick lines.
  The topmost spectrum (TYC~4194~2188) shows absorption due to the Draco Cloud. This
  has been used to predict the expected absorption in the other three
  spectra, as described in the text, and shown as thin lines. Of the
  three remaining spectra, only the bottom two clearly correspond to
  foreground stars. The other spectrum (BD+64~1134) is ambiguous, as
  the expected absorption is small enough that it could simply be
  undetected in these data.}
\end{figure}
\subsection{Distances}
All useful stars (i.e. those with unambiguous identifications as either
foreground or background stars) with classification spectra have been classified by
comparison to the observed set of spectral standards.  The derived
spectral class and the absolute magnitude calibration of Corbally \&
Garrison(1984) were then used to assign an absolute magnitude, and
hence distance, to each star. We have considered three
sources of error in this distance estimate: 1) the error in the Tycho
$V$ magnitude; 2) an assumed classification error of $\pm$ one
spectral class for each star; 3) the intrinsic dispersion (0.7~mag) in
the absolute magnitude relation vs. MK spectral class (Jaschek \&
G\'omez 1998). We have not accounted for extinction, as the
expected overall extinction along typical lines of sight at these
galactic latitudes is small. However, we do expect some extinction of
the background stars due to the intervening IVCs, particularly
because these IVCs are known to be somewhat dusty (HRK88). However,
without detailed multi-band photometry of these stars, the degree of
extinction is unknown, though likely quite
small (Stark {et al.} 1997). Note that the inclusion of extinction 
can only {\it reduce} the estimated upper distance limits. Given that
we now have a spectroscopically identified sample of background stars,
we plan to gather the necessary data to resolve the extinction
problem. In the absence of that data however, we conservatively
include no extinction when estimating distances.
\paragraph{}
For some IVCs classification spectra of some stars have not yet been
acquired. In these cases we have tentatively estimated the distance by
assigning a spectral class on the basis of the $B-V$ colour from the
Tycho catalog. In such cases the distance estimates are often quite
uncertain, as we assign errors to the distance based on the
uncertainty in the colour.

\section{Preliminary Distance Results and Conclusions}
The preliminary distance limits for the our entire target sample of 11
clouds is given in Table 2. While it is apparent that much remains to
be done to refine these measurements, the current data do set some
interesting limits on the distances for these IVCs. In general, all are
more distant than 200pc, and some may be up to several kiloparsecs
distant. Given that all the clouds are at significant Galactic
latitudes, this implies that the bulk of these IVCs are not in the
gaseous disk, but falling in towards it. The DDO IVC distance project
is a continuing effort, and we anticipate refining these distance
estimates in the near future. Moreover, we expect to add many more IVCs to
our sample in an attempt to refine the statistical picture for these
interesting objects.

\begin{table}[!htb]
\begin{center}
\begin{tabular}{ccccc}
\tableline\tableline
Cloud(s) & V$_{LSR}$  & \multicolumn{2}{c}{Distance Limits (pc)}& \# Det's\\
& (km~s$^{-1}$) & Lower & Upper & \# Stars\vspace{0.10cm}\\\hline
G163.9+59.7 & -19.0 & {\it543}$_{{\it-243}}^{{\it+514}}$& {\it1533}$_{{\it-541}}^{{\it+844}}$ & 1 / 5\vspace{0.10cm}\\
G139.6+47.6,G141.1+48.0 & -12.1,-12.9 & 233$_{-116}^{+92}$ &
325$_{-66}^{+93}$ & 8 / 20\vspace{0.10cm}\\
G135.5+51.3  &-47.2 &{\it746}$_{{\it-433}}^{{\it+1747}}$ &1419$_{-362}^{+485}$ &1 / 3\vspace{0.10cm}\\
G149.9+67.4 &-6.3&{\it372}$_{{\it-113}}^{{\it+123}}$ &{\it545}$_{{\it-82}}^{{\it+110}}$ &1 / 3\vspace{0.10cm}\\
G249.0+73.7&-0.6&{\it226}$_{{\it-48}}^{{\it+75}}$ &--&0 / 1\vspace{0.10cm}\\
G124.1+71.6&-11.4&{\it252}$_{{\it-16}}^{{\it+17}}$ &{\it843}$_{{\it-396}}^{{\it+1181}}$ &1 / 2\vspace{0.10cm}\\
G107.4+70.9,G99.3+69.0 &-29.9,-26.6&742$_{-217}^{+324}$
&823$_{-251}^{+393}$ &1 / 8\vspace{0.10cm}\\
G86.5+59.6 &-39.0&604$_{-178}^{+237}$ &--&0 / 10\vspace{0.10cm}\\
G90.0+38.8,G94.8+37.6 &-23.9,-23.3&463$_{-136}^{+192}$ &618$_{-174}^{+243}$&1 / 8\vspace{0.10cm}\\
G81.2+39.2&+3.5&450$_{-132}^{+187}$ &851$_{-276}^{+407}$ & 2 / 5\vspace{0.10cm}\\
G86.0+38.3 &-43.4&921$_{-280}^{+443}$ &2366$_{-793}^{+1192}$&1 / 7\vspace{0.10cm}\\
\tableline\tableline
\end{tabular}
\end{center}
\caption{The preliminary distance results for the DDO IVC distance
  project. Both upper and lower distance limits are given for 9 clouds,
  with only lower limits for 2 clouds. Errors for each distance limit
  are given, computed as described in the main text. Distance limits
  given in italics are from distances based only on $B-V$ colours.} 
\end{table}

\acknowledgments
M.D.G, A.A. and C.R.B. wish to thank the Natural Sciences and Engineering Research
Council of Canada for support through the PGS~A and PGS~B graduate
scholarship programs. We also wish to acknowledge the immense support
this project has received from the Director and staff of the David
Dunlap Observatory. M.D.G. thanks the conference organizers for
financial assistance in journeying from Toronto to Canberra.



\begin{references}
\reference Benjamin, R.A., Venn, K.A., Hiltgen, D.D., Sneden, C. 1996, ApJ,
  464, 836
\reference Burns, C.R. {\it et al.} 1999, in preparation
\reference Clarke, T.E., Mallen-Ornelas,
 G., Sawicki, M., Gladders, M.D., Burns, C.R., Attard, A. 1998, in
 {\it New Perspectives on the Interstellar Medium}, eds. A.R. Taylor,
 T.L. Landecker and G. Joncas, ASP Conference Proceedings, in press (astro-ph/9811066)
\reference Corbally, C.J., Garrison, R.F. 1984, in {\it The MK Process and
    Stellar Classification}, ed. R.F. Garrison, (toronto: DDO), 277
\reference Garcia, B. 1989, Bull. Inf. Cent. de Donnees de Strasbourg, 36, 27
Gladders, M.D., Clarke, T.E., Burns, C.R., Attard, A., Casey, M.P., Hamilton, D., Mallen-Ornelas,
 G., Karr, J.L., Poirier, S.M., Sawicki, M., Barrientos, L.F. and
 Mochnacki, S.M. 1998, \apjl, 507, 161
\reference Hartmann, D., Burton, W.B. 1997, Atlas of Galactic Neutral
  Hydrogen, Cambridge Univ. Press, Cambridge, U.K.
\reference Heiles, C., Reach, W.T., Koo, B. 1988, ApJ, 332, 313 (HRK88)
\reference Hog, E., Baessgen G., Bastian, U., Egret, D., Fabricius, C.,
  Grossman, V., Halbwachs, J.L., Makarov, V.V., Perryman, M.A.C,
  Schwekendiek, P., Wagner, K., Wicenec, A. 1997, A\&A, 323, 57
\reference Jaschek, C., G\'omez, A.E. 1996, A\&A, 330, 619
\reference Kerp, J., Pietz, J., Kalberla, P.M.W., Burton, W. B.,
Egger, R., Freyberg, M. J., Hartmann, D., Mebold, U. 1998, in {\it The
  Local Bubble and Beyond}, eds. D. Breitschwerdt, M.  J.
Freyberg, and J. Truemper, IAU Conference Series vol. 506,
(Springer-Verlag: Berlin), 457
\reference Lilienthal, D., Wennmacher, A., Herbstmeier, U., Mebold, U.
1991, A\&A, 250, 150 
\reference Muller, C.A., Oort, J.H., Raimond, E. 1963, C.R. Acad.
Sci. Paris, 257, 1661
\reference Snowden, S. L., Mebold, U., Hirth, W.,
Herbstmeier, U., Schmitt, J. H. M. M. 1991, Science, 252, 1529
\reference Stark, R., Kalberla, P., G\"usten, R. 1997, A\&A, 317, 907
\reference Wakker, B.P., van Woerden, H. 1997, ARA\&A, 53, 217 (WvW97)

\end{references}
\end{document}